\documentclass[aps,twocolumn]{revtex4-2}
\usepackage [cp1251]{inputenc}
\usepackage{amsmath}
\usepackage{bm}
\usepackage{graphicx}
\usepackage{bm}
\usepackage{epsfig}
\usepackage{epstopdf}
\usepackage{amsmath}
\usepackage{color}

\begin{document}
	
\title{Theory of resonant Raman scattering due to spin-flips of resident charge
	carries \\ and excitons in perovskite semiconductors}
	\author{A. V. Rodina}\email{anna.rodina@mail.ioffe.ru} 
	\author{E. L. Ivchenko}
	\affiliation{Ioffe Institute, 194021 St. Petersburg, Russia}

	\begin{abstract}
We have developed a theory of Raman scattering with single and double spin flips of localized resident electrons and holes as well as nonequilibrium localized excitons in semiconductor perovskite crystals under optical excitation in the resonant exciton region. Scattering mechanisms involving localized excitons, biexcitons and exciton polaritons as intermediate states has been examined, the spin-flip Raman scattering by polaritons being a novel mechanism. The derived equations are presented in the invariant form allowing one for the analysis of the dependence of scattering efficiency on the polarization of the initial and scattered light and on the orientation of the external magnetic field.

		71.35.--y (Excitons and related phenomena), 71.36.+c (Polaritons),  78.30.--j (Infrared and Raman spectra), 81.05.Hd (Other semiconductors) 
		\end{abstract}
	
	\date{October 20, 2022}
	\maketitle

\section{Introduction}

The spin-flip Raman scattering (SFRS) is a powerful method for studying spin interactions in bulk and nanosized semiconductors. The energy shift of the SFRS line in a magnetic field is directly determined by the $g$ factor of the charge carrier or exciton performing the spin flip, and also, in the general case, by the exchange energy of its interaction with other magnetic ions or localized charge carriers. In semiconductors, the SFRS phenomenon was predicted by Yafet in 1961 \cite{Yafet} which was followed by observations of single spin flips of free or localized electrons or holes in semiconductor crystals, and later in semiconductor nanostructures, e.g., \cite{Patel,ToHo1968,ScottReview,SaCa1992,Sirenko1998}, for more detailed references see \cite{Kudlacik2020,Rodina2020}. However, the observation of double or multiple flips of charge carrier spins in the SFRS spectrum is rare, there are few publications on double flips of donor-bound electron spins in bulk CdS \cite{Scott1972,ToHo1968,ScottReview} and CdTe \cite{doubleCdTe} semiconductors as well as double and triple SFRS in ZnTe \cite{OkaCardona}. The theoretical work on SFRS with multiple spin-flips of electrons localized on donors was proposed in \cite{Economou1972}~ and published in the same issue of Physical Review Letters as the first experimental observation \cite{Scott1972}.

In nanostructures, scattering with spin flips of one and two resident localized electrons has been first observed in colloidal CdSe nanoplatelets (NPLs) \cite{Kudlacik2020}. A detailed theory of these processes is presented in Ref.~\cite{Rodina2020}; the scattering mechanisms with different types of intermediate states formed by photoexcitons and resident localized electrons are considered, and expressions for compound matrix elements describing the spin reversal of one or two electrons localized in the same NPL are derived. It is shown that the dependence of SFRS polarization properties on the orientation of the applied magnetic field (in the Voigt or Faraday geometries) allows one to access information about the electron $g$-factor value and its anisotropy in an individual NPL as well as about the orientation of NPLs in the ensemble.

Recently, the SFRS with simultaneous spin reversal of localized electron and hole has been observed in semiconductor perovskites, both in bulk \cite{Kirstein2022,Kalitucha2022} and nanocrystals \cite{Kalitucha_abstract}. Note that the first perovskite CaTiO$_3$ was named in the 19th century by the minerologist Gustav Rose after the Russian count Lev Perovski. Nowadays semiconducting perovskites ABX$_3$ containing organic cations A = MA (methylammonium), FA (formamidinium) or completely inorganic perovskites (e.g., A = Cs; B = Pb, Sn; C = Cl, Br, I), as well as nanocrystals based on them, are actively studied. The interest is due to their unique electronic and optical properties, opening up prospects for their use in numerous  applications \cite{per_app}. To describe the optical properties of perovskites it is important to understand the fine energy structure of excitons caused by crystal symmetry, exchange and spin-orbit interactions of charge carriers \cite{Becker2018,Nestoklon2018,Sercel2019} and to study the coupling of exciton states with light.

In semiconductor lead halide perovskites APbX$_3$, the longitudinal-transverse splitting of exciton states due to the long-range electron-hole exchange interaction can reach several meV. In the reflection spectrum of a bulk sample, such as the CsPbBr$_3$ crystal \cite{Belych2019}, a resonance contour typical for exciton-polaritons \cite{Hopfield1963} is observed. Brillouin scattering of exciton-polaritons has also been detected in bulk CsPbBr$_3$ \cite{Kalitucha2022}. Therefore, in these compounds there are two possible mechanisms for the Raman scattering with spin flips of localized electrons and holes: (1) resonant optical excitation of a localized exciton followed by its exchange interaction with localized electrons and holes, and (2) direct excitation of propagating exciton-polaritons, their scattering on localized carriers, and conversion of polaritons into secondary photons at the sample boundary. In this paper we consider both mechanisms of SFRS in perovskite crystals. The first one resembles the scattering mechanism in CdSe NPLs \cite{Kudlacik2020,Rodina2020} but is characterized by different polarization properties due to the differing symmetry of the objects. It should be noted that in perovskites the simultaneous presence of nonequilibrium, localized and weakly interacting electrons and holes is experimentally confirmed \cite{Belych2019,Kirstein2022,Belych2022}. In general, in the scattering process the initial state can include not only the noninteracting localized electrons and holes but also localized excitons. We will consider both possibilities. The mechanism of SFRS involving the exciton localized as a whole is relevant for the situation in perovsike-based nanocrystals where the weak quantum confinement of the exciton is realized  \cite{Sercel2019}. As for the SFRS mechanism involving exciton-polaritons, it is considered for the first time in this paper. Here we lay the foundation for the theory of SFRS of exciton-polaritons and show how the theory of exciton-polariton transfer \cite{Yuldashev} is generalized with allowance for the exchange interaction of the electron-hole component of a polariton with localized charge carriers.	

The rest of the paper is organized as follows. In Sec. \ref{setup} we describe the symmetry of the electron, hole and exciton states in perovskites both at zero magnetic field (\ref{subA}) and in the presence of an external magnetic field (\ref{subB}) as well as the exchange interaction between resident carriers and an exciton (\ref{subC}). In Sec. \ref{single} we derive compound matrix elements for the single SFRS with photoexcitation of the localized excitons while Sec. \ref{double} treats the double SFRS processes. To this end, we consider the simultaneous spin flip of non-interacting resident electron and hole (Sec.~\ref{IV.A}) and the resonance excitation of a biexciton as the intermediate state in the case of the initial state with a photoexited exciton (Sec.~\ref{IV.B}). Section \ref{excpol} presents a new mechanism of the SFRS related to the direct excitation of the exciton-polaritons. The polarization selection rules for the considered SFRS processes are analyzed in Sec. \ref{VI}, and in Sec. \ref{VII} we make a summary and outline the future work.		
	
\section{Excitons and localized charge carrier states in an external magnetic field} \label{setup}
	
\subsection{Symmetry of band structure states and wave functions of free and localized excitons} \label{subA}
	
Perovskite APbX$_3$ crystals are direct-gap semiconductors; their band structure is inverted in comparison with the III-V and II-VI semiconductors:  the top of the valence band  is formed predominantly by $s$-orbitals of Pb (with slight hybridization of $p$-orbitals of $X$ halogens) and is twice degenerate by the spin projection $s_{h,z}=\pm 1/2$, while the lowest conduction band is formed predominantly by $p$-orbitals of Pb (with slight hybridization of $s$-orbitals of $X$ halogens). As a result of the strong spin-orbit interaction, the sixfold degenerate state in the conduction band is split, with the lowest energy state also twice degenerate by the projection of the total angular momentum of the electron $j_{e,z} =\pm 1/2$ \cite{Becker2018,Kirstein2022}. At room temperature, the Bravais lattice  is the simple cubic (crystal class O$_h$), with band extrema located at the top of the first Brullouin zone, which has a cube shape (${\cal R}$-point, isomorphic to $\Gamma$-point) \cite{Even2015}. As the temperature decreases, the symmetry decreases to tetragonal (crystal class D$_{4h}$) and then to orthorhombic \cite{Steele2020}. We will limit ourselves to a detailed consideration of the cubic phase. 
	
Let us introduce the basis functions of the electron at the bottom of the conduction band and the hole at top of the valence band of the ${\cal R}$-point
\begin{eqnarray} \label{basis}
&& u_{e,\frac12} ({\bm r}) \equiv\ \uparrow_e = - \frac{1}{\sqrt{3}} [ \alpha Z + \beta (X+ {\rm i} Y) ]\: ,\:\\ &&u_{e,-\frac12} ({\bm r}) \equiv\ \downarrow_e =\frac{1}{\sqrt{3}}\: [ \beta Z -
\alpha (X- {\rm i} Y) ] \nonumber\:,\\ && u_{h,\frac12} ({\bm r}) \equiv\ \uparrow_h  =~  \alpha S\;,\; u_{e,-\frac12} ({\bm r}) \equiv\ \downarrow_h =~\beta S \nonumber\:,
\end{eqnarray}
where $S$ is an invariant orbital Bloch function, and $X, Y, Z$ are Bloch functions transformed as coordinates $x,y,z$ by operations of the point group O$_h$; $\alpha$ and $\beta$ are two-component spin columns for spin states with 1/2 and $- 1/2$ projection on the $z$ axis.

	\begin{figure}[h!]
	\centering
	\includegraphics[width=0.75\linewidth]{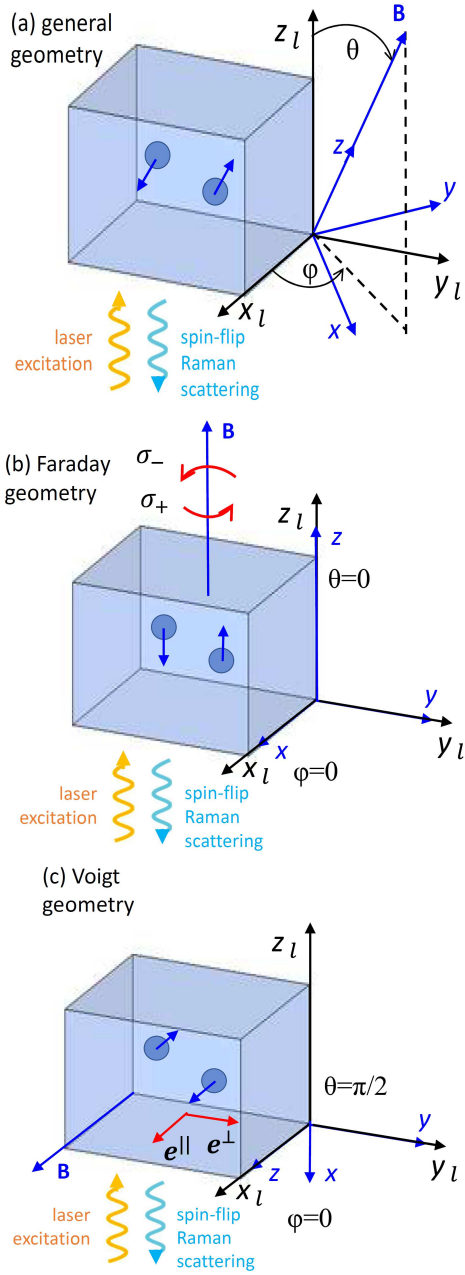}
	\caption{  (a) The schematic representation of the light scattering geometry. The balls illustrate two resident localized charged carriers with their spins $\uparrow_1$ and $\downarrow_2$ oriented along and counter to the magnetic field ${\bm B}$. The axes $x_l, y_l, z_l$ and $x,y,z$ represent the laboratory and field-related coordinate systems. (b) The Faraday geometry. (c) The Voigt geometry. 	\label{Fig1}  } 
\end{figure}

In the cubic perovskite phase, the model of the electronic band structure including only the lowest conduction band and the upper valence band is isotropic and the choice of coordinate frame axes is arbitrary. To describe the spin states of the resident carriers and excitons, it is convenient to choose the axes $x,y,z$ with $z$ oriented along to the external magnetic field ${\bm B}$. In addition, in order to consider the spin-flip Raman scattering processes with the arbitrary orientation of the magnetic field and light propagation direction, we introduce a second, laboratory, coordinate frame  $x_l, y_l, z_l$, in which the axis $z_l$ is directed along the normal to the sample surface. The orientation of the axis $z \parallel {\bm B}$  in the laboratory frame is determined by the polar angles $\theta$ and $\varphi$ as shown in Fig. \ref{Fig1}(a). For simplicity, we will assume  that the incident light propagates along the normal to the surface of the substrate in the positive direction of the $z_{l}$ axis of the laboratory frame, and the scattered light is collected along or backward along this axis. Figure \ref{Fig1}(a) and Fig. \ref{Fig1}(b) show the cases of Faraday ($\theta=0$) and Voigt ($\theta =\pi/2$) geometry, respectively. 

The four-fold degeneracy of the band-edge exciton level is partially removed due to the exchange interaction between the electron and hole bound in the exciton, which can be represented as
\begin{equation} \label{e-e-h}
	{\cal H}_{e \mbox{-}h }= J_{e h} {\bm \sigma}_{e} \cdot {\bm \sigma}_h  \, .
\end{equation}
Here $J_{e h}$ is the energy constant of the exchange interaction, 
${\bm \sigma}_h$ $-$ three-component pseudovectors whose projections are the Pauli matrices acting on the spin states $\uparrow_h$ and $\downarrow_h$. As for the operators $\sigma_{e,l}$ $(l = x,y,z)$, for convenience they are defined so that the eigenfunctions of the operator $\sigma_{e,z}$ are the basis functions $\uparrow_e, \downarrow_e$ in \eqref{basis}, not the spin columns $\alpha$ and $\beta$. With this choice the hole spin and total angular momentum operators of the electron can be represented as ${\bm s}_h={\bm \sigma}_h/2$ and ${\bm j}_e={\bm \sigma}_e/2$, respectively. 
		
The constant $J_{e h}$ includes the contributions of the short- and long-range exchange interaction. For the first mechanism, the scalar product of the vector matrices ${\bm \sigma}_{e}$ and ${\bm \sigma}_h$ in the right-hand side (\ref{e-e-h}) follows from the symmetry of the short-range contact potential. For the second mechanism, this type of interaction is applicable for isotropic localization of the exciton when the two-part envelope $\Phi_{\rm exc}({\bm r}_e, {\bm r}_h)$ (${\bm r}_e$ and ${\bm r}_h$ are coordinates of the electron and hole in the exciton) is invariant with respect to the coordinate frame rotations. For an anisotropic localization of the exciton, the exchange interaction operator has a more complex form \cite{Goupalov1998}. We will assume that the additional anisotropy-induced splittings of the exciton level  are small compared to the damping of the exciton $\hbar \Gamma$ and the formula (\ref{e-e-h}) is applicable.

The exchange interaction \eqref{e-e-h} leads to the formation of triplet and singlet exciton states. The wave function of a singlet exciton with zero total momentum $J=0$ has the form
\begin{equation}\label{basis0}
\Psi_{0,0}(\bm r_e,\bm r_h) = \frac{1}{\sqrt{2}} (\uparrow_e \downarrow_h - \downarrow_e \uparrow_h)  \Phi_{\rm exc}(\bm r_e,\bm r_h)\:.
\end{equation}
We choose the basis  wave functions of a triplet exciton with total momentum $J=1$  as
\begin{equation}\label{basis1}
\Psi_{1,j}(\bm r_e,\bm r_h) = \Phi_{\rm exc}(\bm r_e,\bm r_h)\ v_j  \:.
\end{equation}
Here	$v_{j}$ $(j=x,y,z)$ are the two-particle Bloch functions
\begin{eqnarray} \label{basis2}
v_x &=& \frac{1}{\sqrt{2}} ( - \uparrow_e \uparrow_h + \downarrow_e \downarrow_h) \:\nonumber ,\\ v_y &=&\frac{\rm i}{\sqrt{2}} (\uparrow_e \uparrow_h + \downarrow_e \downarrow_h) \:,\\ v_z &=& \frac{1}{\sqrt{2}}  (\uparrow_e \downarrow_h + \downarrow_e \uparrow_h) \nonumber\:,
\end{eqnarray}
transformed as coordinates $x,y,z$. The two-particle envelope  wave function $\Phi_{\rm exc}(\bm r_e,\bm r_h)$ describes the state of a free exciton as well as an exciton localized as a whole on fluctuations of potential  or at a  defect in a bulk crystal. It can be represented as
\begin{equation} \label{Phiexc}
\Phi_{\rm exc}({\bm r}_e, {\bm r}_h) = f({\bm r}_e- {\bm r}_h) F({\bm R})\, ,
\end{equation}
where the function $f({\bm r})$ describes the relative motion of the electron and hole, while the motion of the exciton center of mass, ${\bm R} = (m_e{\bm r}_e+m_h{\bm r}_h )/M$ ($M=m_e + m_h$ $-$ the translational mass of the exciton)  is described by the function $F({\bm R})$.  In the case of a free mobile exciton, the motion of the center of mass and thus the quantum excitation is characterized by the wave vector ${\bm k}$, so that 
\begin{equation} \label{freek}
F({\bm R}) \equiv	F_{\bm k}({\bm R}) = \frac{{\rm e}^{{\rm i} {\bm k}{\bm R}}}{\sqrt{V}} \:,
\end{equation}
where $V$ is the normalization volume. For localized states of excitons, the envelope function $F({\bm R})$ can generally be written in the form of a Fourier function integral expansion (\ref{freek}).

In the dipole approximation, the singlet exciton $\Psi_{0,0}$ does not interact with light, it is the so-called ``dark'' exciton.  The matrix elements of optical excitation of triplet (``bright'') excitons $\Psi_{1,j}$ have the form 
\begin{equation} \label{meexc}
M_{j}^{(\rm abs)}({\bm e}^0) {\cal E}^0 = \sqrt{\frac23} d_{\rm cv} {\cal I}_{\Phi} {\cal E}^0 e^0_j \:.
\end{equation}
Here ${\cal E}^0$ and ${\bm e}^0$ are the amplitude and unit polarization vector of incident light, 
\begin{equation} \label{calPHY}
{\cal I}_{\Phi} = \int \Phi_{\rm exc}({\bm r}, {\bm r}) d {\bm r}\:,
\end{equation}
$d_{\rm cv}$ is the interband matrix element of the dipole momentum operator $e \langle X| x | S \rangle = e \langle Y | y | S \rangle = {e \langle Z| z | S \rangle}$ calculated between the Bloch functions at the ${\cal R}$-point of the Brullouin zone. In the general case, light excites an exciton 
\begin{equation} \label{exciton}
	\Psi = e^0_{x} \Psi_{1,x} + e^0_{y} \Psi_{1,y}+ e^0_z \Psi_{1,z}\:.
\end{equation} 
For simplicity, we will assume that the incident light propagates along the normal to the surface of the substrate in the positive direction of the $z_{l}$ axis of the laboratory frame, and the scattered light is collected along or backward along this axis.

For the emission matrix element, to within a multiplier, we have 
\begin{equation} \label{meem}
M_j^{(\rm em)}({\bm e}) = M_j^{(\rm abs)*}({\bm e}) = \sqrt{\frac23} d_{\rm cv}{\cal I}_{\Phi}  e^*_j \:.
\end{equation}
The factor ${\cal I}_{\Phi}$ is an enhancement factor of the exciton oscillator strength due to localization and, for a localized exciton with the localization length $L_{\rm exc}$, is proportional to
$L_{\rm exc}^{3/2}$ in a bulk semiconductor and $L_{\rm exc}$ in two dimensions. The giant oscillator strength of a localized exciton proportional to $L_{\rm exc}^3$ was predicted by Rashba and Gurgenishvili 60 years ago  \cite{Rashba}.
\subsection{Localized charge carriers in a magnetic field} \label{subB}
The spin splittings of the resident electron and hole in the magnetic field ${\bm B}$ are controlled by the effective $g$ factors (Land\'e factors), $g_e$ and $g_h$, and are described by Hamiltonians
\begin{eqnarray} \label{gfactor}
{\cal H}_e = \frac{1}{2} g_e \mu_B {\bm \sigma}^r_e \cdot {\bm B}\:,\:{\cal H}_h = \frac{1}{2} g_h \mu_B {\bm \sigma}^r_h \cdot {\bm B}\:.
\end{eqnarray}
Here $\mu_B$ is the Bohr magneton, the operators ${\bm \sigma}^r_e$, ${\bm \sigma}^r_h$ act on the basis functions \eqref{basis} entering the wave functions of resident electrons and holes
\begin{eqnarray}  \label{loceh}
&&{\psi}_{e, 1/2} = \phi_e({\bm r} - {\bm r}^0_e) \uparrow_e\:, {\psi}_{e,-1/2} = \phi_e({\bm r} - {\bm r}^0_e) \downarrow_e\:, \\
&&{\psi}_{h,1/2} = \phi_h({\bm r}- {\bm r}^0_h) \uparrow_h\:, {\psi}_{h,-1/2} = \phi_h({\bm r} - {\bm r}^0_h) \downarrow_h\:, \nonumber
\end{eqnarray}
where $\phi_e, \phi_h$ are the envelopes of localized states, ${\bm r}_{e}^0, {\bm r}_h^0$ are the spatial positions of the defects at which the charge carriers are localized. In the coordinate frame $x,y,z$ the scalar products in (\ref{gfactor}) can be replaced by $\sigma^r_{e,z} B$ and $\sigma^r_{h,z} B$ and the eigenstates of the Zeeman Hamiltonians (\ref{gfactor}) are functions (\ref{loceh}). In the following we make no difference between $g$ factor values of  localized and exciton-bound particles. 

The recently published paper \cite{Kirstein2022} presents experimental data and results of theoretical calculations (in the density functional method and in the second order ${\bm k}\cdot{\bm p}$ perturbation theory) of the electron and hole $g$ factors in bulk lead halide perovskites as functions of  the bandgap width $E_g$. It is shown that the $g_e$ values turn out to be positive in all the studied materials, whereas the $g_h$ values can be either negative (at $E_g < 1.8$ eV) or positive (at $E_g > 1.8$ eV). For definiteness, we will further set $g_e > |g_h|$. Moreover, in what follows we will consider only magnetic fields ${\bm B}$ fulfilling the inequalities
\begin{equation}
 g_e \mu_B B, |g_h| \mu_B B \ll \hbar \Gamma \, . 
 \label{gehG}
 \end{equation}
 	
\subsection{Exchange interaction between exciton and localized charge carriers}
\label{subC}
We assume here that the resident carriers with envelope wave functions $\phi_e({\bm r})$ and $\phi_h({\bm r})$ are localized far enough apart to neglect the exchange interaction between them in comparison with the Zeeman energies. The exchange interaction between the localized electron or hole with the exciton is realized through the exchange interaction with the electron or hole bound in the exciton, respectively, which has the form 
\begin{equation} \label{exchange2}
	H_{\rm exch} = \tilde{J}_{ee} \Omega_0 {\bm \sigma}_e {\bm \sigma}^r_e \delta({\bm R} - {\bm r}_e^0) + \tilde{J}_{hh} \Omega_0 {\bm \sigma}_h {\bm \sigma}^r_h \delta({\bm R} - {\bm r}_h^0)\:.
\end{equation}
Here the volume of the unit cell $\Omega_0$ is introduced in order to have for the coefficients $\tilde{J}_{ee}, \tilde{J}_{hh}$ the energy dimension. We neglect the exchange interaction between the dissimilar localized and bound-to-exciton particles as compared with the interactions \eqref{e-e-h} and \eqref{exchange2}.  The energy parameters $\tilde{J}_{ee}$ and $\tilde{J}_{h h}$ depend on the exciton Bohr radius and the localization radius of the localized electron or hole and can be found according to the procedure described for the exciton and localized electrons in \cite{Rodina2020}. 

For a localized exciton with the center-of-mass envelope $F({\bm R})$, the operator (\ref{exchange2}) is transformed to the sum of spin operators
\begin{equation} \label{e-e-e}
{\cal H}_{e \mbox{-}e }=  J_{ee} {\bm \sigma}_{e} \cdot {\bm \sigma}^r_e \:,\:{\cal H}_{h\mbox{-}h}= J_{hh} {\bm \sigma}_{h} \cdot {\bm \sigma}^r_h \:,
\end{equation}
where
\[
J_{ee} = \tilde{J}_{ee} \Omega_0 F^2({\bm r}_e^0)\:,\:J_{hh} = \tilde{J}_{hh} \Omega_0 F^2({\bm r}_h^0)\:.
\]

The matrix elements of the Pauli matrix operators between the states of the triplet exciton (\ref{basis1}) and the singlet exciton (\ref{basis0}) are
\begin{eqnarray} \label{pauli_exc}
&&	\langle \Psi_{1,j} | \sigma_{e, k} | \Psi_{1,l} \rangle =\langle \Psi_{1,j} | \sigma_{h, k} | \Psi_{1,l} \rangle = {\rm i} {\rm e}_{jkl} \, ,  \\ 
&&	\langle \Psi_{0,0} | \sigma_{e, k} | \Psi_{1,l} \rangle = -\langle \Psi_{0,0} | \sigma_{h, k} | \Psi_{1,l} \rangle = \delta_{kl} \, , \nonumber
\end{eqnarray}
with $j,k,l=x,y,z$, ${\rm e}_{jkl}$ being the unit antisymmetric pseudotensor of the third rank and $\delta_{kl}$ being the unit symmetric tensor of the second rank. 
\section{Light scattering with a single spin flip: localized excitons} \label{single}
\subsection{Intensity and compound matrix element of the scattering process} \label{singleA}
Here we consider inelastic light-scattering with a single spin flip of the resident carrier under photoexcitation of a localized exciton. For definiteness, we assume the electron and hole $g$ factors to be positive and focus on scattering in the Stokes region of the spectrum, $\omega < \omega_0$, where $\omega_0$ and $\omega$ are the frequencies of incident and scattered light, respectively. Then the intensity of light undergoing the single spin-flip scattering has the form
\begin{eqnarray}  \label{intensity10}
	&&\hspace{- 2 mm} I^{(1e)}_{+} \propto |V_{f,i}^{(1e)} |^2 \delta (\hbar \omega_0 - \hbar \omega - g_e \mu_{\rm B} B ) f_{\downarrow_e} ,\\
	&&\hspace{- 2 mm} I^{(1h)}_{+}  \propto |V_{f,i}^{(1h)} |^2 \delta( \hbar \omega_0 - \hbar \omega - g_h \mu_{\rm B} B ) f_{\downarrow_h}\:, \nonumber
\end{eqnarray}
where $V_{f,i}^{(1e)}$ and $V_{f,i}^{(1h)}$ are the compound matrix elements of the scattering from the initial state of a resident electron with spin $i =\ \downarrow_e$ or resident hole with spin $i =\ \downarrow_h$ to the final state $f =\ \uparrow_e$ or $f =\ \uparrow_h$. Such processes can occur in crystals containing any number of resident carriers, taking into account the interaction of the photoexcited exciton interacts with only one of them. The expressions (\ref{intensity10}) include the occupation of the initial electron (hole) state, which at a fixed temperature $T$ are defined by Fermi functions 
\begin{equation}
f_{\downarrow_{e(h)}} = \left[ 1+ \exp (-g_{e(h)}\mu_{\rm B}B/k_{\rm B}T) \right]^{-1} \, , 
\end{equation}
where $k_{\rm B}$ is the Boltzmann constant. For the scattering intensities $I^{(1e)}_{-}$ and $I^{(1h)}_{-}$ into the anti-Stokes region, the minus sign in front of the $g$ factors in the $\delta$-functions in Eqs.~\eqref{intensity10} should be changed to plus, and the occupation 
$f_{\downarrow_{e(h)}}$ replaced by the occupation  $f_{\uparrow_{e(h)}}=1-f_{\downarrow_{e(h)}}$ for the initial spin-up state of the resident electron (hole). Definitely, the initial and final states $i, f$ also include the incident and scattered photons with the energies $\hbar \omega_0$, $\hbar \omega$ and the unit polarization vectors ${\bm e}^0, {\bm e}$, respectively.

\subsection{Three-particle intermediate state ``exciton plus localized electron or hole''} \label{singleB}
The spin structure of the intermediate state formed by a photoexcited exciton and a resident carrier depends on the ratio between the energy $J_{ee}$ or $J_{hh}$ of their interaction and the electron-hole exchange energy $J_{eh}$.

In general, the intermediate spin states are characterized by the projections $\pm 1/2$ and $\pm 3/2$ of total spin on the magnetic field. In the limiting case of a strong exchange interaction between a resident carrier and a similar carrier in an exciton, e.g., $|J_{ee}| \gg |J_{eh}|$ for the resident electron, the two electrons after photoexcitation form a singlet and a triplet state with total spin 0 and 1, respectively.  In the opposite case of a strong exchange interaction between the electron and hole in the exciton, $ |J_{eh}| \gg |J_{ee}|, |J_{hh}|$, the \eqref{basis0}, \eqref{basis2} play role of the intermediate states. In this case, similarly to the situation studied in Ref.~\cite{Rodina2020}, there exist direct and indirect channels  for excitation and recombination of excitons weakly interacting with localized resident charge carriers. We remind that, e.g., in the indirect recombination channel, the electron (or hole) in the exciton recombines with the resident hole (or electron) and the remaining photoexcited charge carrier takes the place of the latter. The probabilities of such processes contain indirect overlap integrals ${\cal I}_{r} =  \iint \phi_{e(h)}({\bm r})\phi_{e(h)}({\bm r}')\Phi_{\rm exc}({\bm r}, {\bm r}') d {\bm r} d {\bm r}'$ which are usually significantly smaller than ${\cal I}_{\Phi}$ controlling the direct exciton excitation and recombination. 

To further simplify the consideration, we limit ourselves here to weak overlap and weak exchange interaction of excitons with resident carriers and take into account only the direct excitation and recombination channels of excitons. In this case, the matrix element of the single spin-flip process reads
	\begin{equation} \label{MEsingle}
		V_{f,i}^{(1e(1h))} =   \sum\limits_{n' n} \frac{ M^{({\rm em})}_{f,n'}({\bm e}) \Delta^{e(h)}_{n', n} M^{({\rm abs})}_{n,i}({\bm e}^0) }{ ( E_{1} - \hbar \omega_0 - {\rm i} \hbar \Gamma_1) ^2}  \:.
	\end{equation}
Here, the three-particle intermediate states $n$ and $n'$ include photoexcited bright excitons $\Psi_{1,j}, \Psi_{1,j'}$  ($j,j'=x,y,z$), differ in the spin direction of the resident carrier and, for the Stokes process, can be written as 
\begin{equation} \label{nn'}
n = \Psi_{1,j}	,  \downarrow_{e(h)}~\mbox{and}~n' = \Psi_{1,j'},  \uparrow_{e(h)}\:.
	\end{equation}
Other notations in Eq.~(\ref{MEsingle}) are $E_1$ for the excitation energy of the bright exciton,  the decay parameter $\Gamma_1$ for the radiative and nonradiative recombination of the exciton and the finite lifetime of the resident carrier in the localized state. The matrix elements of light absorption and emission, $M^{({\rm abs})}_{n,i}( {\bm e}^0 )$ and $M^{({\rm em})}_{f,n}({\bm e})$, are defined according to (\ref{meexc}) and (\ref{meem}). The matrix element of the exchange interaction between the resident electron and the exciton-bound electron describes the spin flip of the resident carrier as follows
\begin{eqnarray}  \label{Dj}
\Delta^{e}_{n', n} &=& \frac{J_{ee}}{2} \langle \Psi_{1,j'} |\sigma_{e,-} | \Psi_{1,j} \rangle \langle \uparrow_e | \sigma_{e,+}^r |\downarrow_e \rangle  \nonumber \\
 &=& J_{ee} \langle \Psi_{1,j'} |\sigma_{e,-} | \Psi_{1,j} \rangle \:, 
\end{eqnarray} 
where $\sigma_{\pm} = \sigma_x \pm {\rm i} \sigma_y$. For the interaction between the resident and exciton-bound holes, the index $e$ should be changed to $h$.

Using the first equation~(\ref{pauli_exc}) and Eqs.~\eqref{meexc},\eqref{meem}  we find the matrix element of the single SFRS 
\begin{equation} \label{CSCT1}
	V_{f,i}^{(1e(1h))}=  - \frac{2{\rm i}}{3} \frac{ J_{ee(hh)} d_{\rm cv}^2 {\cal I}_{\Phi}^2 {\cal E}^0 }{(E_1  - \hbar \omega_0 - {\rm i} \hbar \Gamma_{1})^2} ({\bm e}^*\times{\bm e}^0) \cdot ({\bm o}_x - {\rm i} {\bm o}_y).
\end{equation}
Here ${\bm o}_x$, ${\bm o}_x$ are the unit vectors (orts) along the axes $x,y$, perpendicular to the magnetic field direction ${\bm B}$ (Fig. \ref{Fig1}). For the anti-Stokes process $\uparrow\ \to\ \downarrow$, the vector ${\bm o}_{-} = {\bm o}_x - {\rm i} {\bm o}_y$ should be replaced by ${\bm o}_{+} = {\bm o}_x + {\rm i} {\bm o}_y$. 

The polarization dependence of the intensity $I^{(1e(1h))} \propto |V_{f,i}^{(1e(1h))}|^2$ is analyzed in Sect. \ref{VI}. Here we just explain the symmetry aspect of the scalar product in Eq.~(\ref{CSCT1}). Let us consider an auxiliary problem of the spin reversal $\downarrow\ \to\ \uparrow$ caused by a perturbation ${\bm \sigma} \cdot {\bm h}$, where ${\bm h}$ is the amplitude of the effective pseudovector forse. The matrix element of such a transition is equal to
\begin{eqnarray} \label{M1}
&&\uparrow^{\dag} \left[ \begin{array}{cc} h_z & h_x - {\rm i} h_y \\ h_x + {\rm i} h_y & - h_z\end{array}\right] \downarrow\ = {\bm h}\cdot ({\bm o}_x - {\rm i} {\bm o}_y)\:. 
\end{eqnarray}
In the spin-flip light scattering, the role of the pseudovector ${\bm h}$ is played by the vector product ${\bm e}^*\times{\bm e}^0 $ in (\ref{CSCT1}). Note, that ${\bm b}\cdot ({\bm o}_x \pm {\rm i} {\bm o}_y)=0$, where ${\bm b}= {\bm B}/B$ is a unit vector along magnetic field direction. 
\section{Double spin-flip Raman scattering: localized excitons \label{double}}
In this section we will consider two mechanisms of double light scattering processes with simultaneous reversals of electron and hole spins. In the first mechanism, section \ref{IV.A}, the spin flips are experienced by a resident electron and hole localized in a perovskite crystal; in the intermediate resonant state an exciton localized in the crystal on the structure defect is added to the electron and hole. In the second mechanism, section \ref{IV.B}, there is a nonequilibrium localized exciton in the sample in the initial and final states, and the role of the intermediate state is played by a biexciton.
\subsection{Localized electron, hole, and exciton} \label{IV.A}
Let the sample simultaneously contain localized resident electron and hole. 
Four possible initial states of localized electrons and holes determine four double spin reversal processes: ``$++$'' and ``$+- $'' from the initial state $\downarrow_e \downarrow_h$ to the final state $\uparrow_e \uparrow_h$ and from the initial state $\downarrow_e \uparrow_h$ to the final state $\uparrow_e \downarrow_h$ (Stokes shift for $g_e > g_h > 0$) and anti-Stokes transitions $\uparrow_e \uparrow_h$ $\to$ $\downarrow_e \downarrow_h$, $\uparrow_e \downarrow_h$ $\to$ $\downarrow_e \uparrow_h$, denoted by ``$-$\hspace{0.3 mm}$-$'' and ``$-+$'' respectively. 

For the $++$, $+-$ processes, the cross sections of the double scattering are proportional to
\begin{eqnarray}  \label{intensity20}
	&&\hspace{- 2 mm} I_{++}^{(1e,1h)} \propto |V_{f,i}^{++} |^2 \delta[\hbar \omega_0 - \hbar \omega - (g_e + g_h) \mu_{\rm B} B] f_{{\downarrow}_e} f_{{\downarrow}_h},\nonumber\\
	&&\hspace{- 2 mm} I_{+-}^{(1e,1h)} \propto |V_{f,i}^{+-} |^2 \delta[\hbar \omega_0 - \hbar \omega - (g_e - g_h) \mu_{\rm B} B] f_{{\downarrow}_e} f_{{\uparrow}_h} , \nonumber\\
\end{eqnarray}
where $V_{f,i}^{++}$, $V_{f,i}^{+-}$ are the compound matrix elements calculated in the fourth order of perturbation theory. Each comprises two terms
\begin{eqnarray}  \label{VVVV}
&&V_{f,i}^{++} = V_{f,i}(h_+ e_+) + V_{f,i}(e_+ h_+)\:, \\ && V_{f,i}^{+-} = V_{f,i}(h_- e_+) + V_{f,i}(e_+ h_-)\:, \nonumber
\end{eqnarray} 
where the symbols $h_{\pm} e_+$ ($ e_+ h_{\pm}$) denote the process in which first the electron (hole) experiences a spin flip and then the hole (electron) does. 

The matrix elements entering Eqs.~(\ref{VVVV}) are found to be
\begin{widetext}
\begin{eqnarray} \label{MEdouble}
&&	V_{f,i}(h_{\pm} e_+) = {\cal E}^0\ \sum\limits_{n'' n' n} \frac{ M^{({\rm em})}_{f,n''}({\bm e}) \Delta^h_{n'', n'} \Delta^e_{n', n} M^{({\rm abs})}_{n,i}({\bm e}^0) }{ ( E_{J'} - \hbar \omega_0 - {\rm i} \hbar \Gamma_{J'})  ( E_1 - \hbar \omega_0 - {\rm i} \hbar \Gamma_1) ^2}  \:, \\ && V_{f,i}(e_+h_{\pm}) = {\cal E}^0\ \sum\limits_{n'' n' n} \frac{ M^{({\rm em})}_{f,n''}({\bm e}) \Delta^e_{n'', n'} \Delta^h_{n', n} M^{({\rm abs})}_{n,i}({\bm e}^0) }{ ( E_{J'} - \hbar \omega_0 - {\rm i} \hbar \Gamma_{J'})  ( E_1 - \hbar \omega_0 - {\rm i} \hbar \Gamma_1) ^2}\:,\nonumber
\end{eqnarray}
\end{widetext}
where $J'$ is the angular momentum of the exciton in the intermediate state $n'$. Compared to the single spin-flip process, the matrix elements (\ref{MEdouble}) involve three intermediate states $n, n', n''$ of the complex ``localized photoexciton + localized resident electron and hole''. 

The states $n$ and $n''$ include the bright excitons $\Psi_{j}, \Psi_{j''}$ ($j,j''=x,y,z$) and can be written as
\begin{equation} \label{nn''}
n = \Psi_{1,j}, \downarrow_e \downarrow_h~~\mbox{and}~~n'' = \Psi_{1,j''}	,  \uparrow_e \uparrow_h\:.
	\end{equation}
As for the $n'$ states, there are eight of them. In particular, for the scattering processes $h_+ e_+$ and $e_+ h_+$ these states are
\begin{equation} \label{n'}
n' = \Psi_{J',j'},  \uparrow_e \downarrow_h~~\mbox{and}~~n' = \Psi_{J',j'},  \downarrow_e \uparrow_h\:,
	\end{equation}
including the dark exciton state $\Psi_{0,0}$.
\subsubsection{The $h_+ e_+$ and $e_+ h_+$ processes}
The calculation shows that the processes $h_+ e_+$ and $e_+ h_+$ make identical contributions, so that it suffices to calculate the contribution of the first four intermediate states to (\ref{n'}), and allowance for the second four by doubling $V_{f,i}(h_+ e_+)$. 

In addition to the matrix elements of the exchange interaction (\ref{Dj}), it is necessary to calculate similar matrix elements between the bright states $\Psi_{1,j}$ ($j=x,y,z$) and the dark exciton state, $\Psi_{0,0}$, using the second line of Eq.~(\ref{pauli_exc}). Omitting intermediate transformations, we present the result
\begin{eqnarray} \label{MEdouble2}
&& \mbox{} \hspace{1.5 cm} V_{f,i}^{++}= \frac43 \frac{d_{cv}^2 J_{ee} J_{hh} {\cal I}_{\phi}^2 {\cal E}^0}{(E_1 - \hbar \omega - {\rm i} \hbar \Gamma_1)^2} \\&&\mbox{} \hspace{0.5 cm}\times \left( \frac{R_1(h_+e_+) }{E_1 - \hbar \omega - {\rm i} \hbar \Gamma_1} + \frac{R_0(h_+e_+)}{E_0 - \hbar \omega - {\rm i} \hbar \Gamma_0} \right)\:, \nonumber \\
&&   R_{1}(h_+e_+) = 
R_{0}(h_+e_+) =   \nonumber \\
&& \mbox{} \hspace{1.5 cm}- [{\bm e}^* \cdot ({\bm o}_x - {\rm i} {\bm o}_y)][{\bm e}^0 \cdot ({\bm o}_x - {\rm i} {\bm o}_y)]
\:. \label{RN}
\end{eqnarray}
For the anti-Stokes process, the matrix element $V_{f, i}^{--}$ contains the invariant $$R_{1}(h_-e_-)= R_{0}(h_-e_-) = [{\bm e}^* \cdot ({\bm o}_x + {\rm i} {\bm o}_y)][{\bm e}^0 \cdot ({\bm o}_x + {\rm i} {\bm o}_y)].$$ 

It is interesting to compare this result with the matrix element of the double-spin-flip process involving two electrons (or two holes) localized in the sample. The matrix element $V_{f,i}(e_+e_+)$ can also be represented in the form (\ref{MEdouble2}) where $J_{ee} J_{hh}$ is replaced by the product $J_{ee,1} J_{ee,2}$ (of the exchange energies of the exciton with the first and second resident electrons) and the sign in the brackets is reversed, so that
\begin{equation} \label{Ree123}
R_1(e_+e_+) = - R_0(e_+e_+) = R_1(h_+e_+)\:.
\end{equation}
Therefore, the energy denominators $E_J - \hbar \omega - {\rm i} \hbar \Gamma_J$ ($J=0, 1$) enter  (\ref{MEdouble2}) with the same sign for $h_+e_+$ scattering, and with opposite signs for $e_+e_+$ or $h_+h_+$ scattering. Thus, in the case of the small exchange splitting of the exciton level, $\Delta_{\rm exc} = |E_1-E_0| \ll \hbar \Gamma_1$ and comparable exciton broadening $\hbar \Gamma_0$ and $\hbar \Gamma_1$, the $h_+e_+$ process goes much more efficiently compared to the $e_+e_+$ or $h_+h_+$ process. 

\subsubsection{The $h_- e_+$ and $e_+ h_-$ processes}
The contributions to the compound matrix element $V_{f,i}^{+-}$ from the processes $h_- e_+$ and $e_+ h_-$ do not coincide. Their sum can be represented by Eq.~ (\ref{MEdouble2}) where $R_J(h_+e_+)$ is replaced by $R_J^{(+-)} = [R_J(h_-e_+)+R_J(e_+h_-)]/2$ ( $J=0,1$) where 
\begin{eqnarray} 
	&& R_1^{(+-)} = 
	 \left[ ({\bm e}^* \cdot {\bm e}^0) +  ({\bm e}^* \cdot {\bm b})({\bm e}^0 \cdot {\bm b}) \right]\:,     \\                 \label{R1(h-e+)}
	 	&&  R_0^{(+-)} = 
	 - \left[ ({\bm e}^* \cdot {\bm e}^0) -  ({\bm e}^* \cdot {\bm b})({\bm e}^0 \cdot {\bm b}) \right]  \:.  \label{R0(h-e+)}
\end{eqnarray}

It should be noted that the $++$ and $--$ processes change the value of the total projection of the resident carrier spin $m = s_{h,z} + j_{e,z}$ on the magnetic field direction by $\Delta m= 2$ and $-2$, respectively, while for the $+-$ and $-+$ processes $\Delta m =0$. In an isotropic medium, the the value $|\Delta m| =2$ is a maximum possible for the SFRS. Therefore, for processes involving three or larger number of resident carriers, processes  $+++$ or $---$ etc. are forbidden but the processes of the kind $+-+$ and $-+-$ etc. are still allowed. 
\subsection{Biexciton as an intermediate state} \label{IV.B}
In this section we ignore resident electrons and holes separately located and consider a Stokes scattering process in which the initial state $i$ involves an incident photon of the energy $\hbar \omega_0$ and the unit polarization vector ${\bm e}^0$ and a nonequilibrium localized exciton $\Phi_{\rm exc}({\bm r}_e, {\bm r}_h)$$\downarrow_e \downarrow_h$ in the spin state with the spin $z$-component $m = s_{e,z} + j_{h,z} = -1$;  the final state $f$ comprises a scattered photon of the energy $\hbar \omega$ and the polarization ${\bm e}$ and a localized exciton $\uparrow_e \uparrow_h \Phi_{\rm exc}({\bm r}_e, {\bm r}_h)$ with the spin component $m = +1$. The intermediate states $n$ are the states of the biexciton $XX$ formed by the initially and secondary  photoexcited excitons. The energy conservation law $\hbar (\omega_0 - \omega) = (g_e + g_h) \mu_B B$ in such a process is described by the first delta-function in Eqs.~(\ref{intensity20}). However, its intensity $I_{++}(XX)$ is proportional to the exciton state occupancy $f_{\downarrow_e \downarrow_h}$ instead of the product $f_{\downarrow_e} f_{\downarrow_h}$. The compound matrix element of the process can be found in the second order of perturbation theory as 
	\begin{eqnarray} \label{MEexciton}
		&&	V_{f,i}^{++}(XX) = {\cal E}^0\ \sum\limits_{n} \frac{ M^{({\rm em})}_{f,n}({\bm e})  M^{({\rm abs})}_{n,i}({\bm e}^0) }{ ( E_{n,{\rm bi}} - \hbar \omega_0 - {\rm i} \hbar \Gamma_{n,{\rm bi}}) }  \:.
	\end{eqnarray}
We restrict ourselves to only one biexciton state $n$ where the spins of two electrons and the spins of two holes form singlets and the envelope function of the biexciton, $\Phi_{\rm biexc}({\bm r}_{e1}, {\bm r}_{e2},{\bm r}_{h1},{\bm r}_{h2})$, is symmetric with respect to the coordinate
exchanges ${\bm r}_{e1} \leftrightarrow {\bm r}_{e2}$ and ${\bm r}_{h1} \leftrightarrow {\bm r}_{h2}$. Triplet biexciton states in bulk crystals are less stable, they are characterized by a large damping and, as a consequence, give a small contribution to the light scattering. The matrix elements of excitation and recombination of the singlet biexciton from and into the exciton states $\Psi_{1,j}$ ($j=x,y,z$) have the form of Eqs.~\eqref{meexc},\eqref{meem} where the enhancement factor $I_{\rm \Phi}$ must be replaced by
$$
I_{XX} =  \iiint \Phi_{\rm biexc}({\bm r}, {\bm r}_{e},{\bm r},{\bm r}_{h})\Phi_{\rm exc}( {\bm r}_{e},{\bm r}_{h}) d {\bm r} d {\bm r}_e d {\bm r}_h \, .
$$

Similarly to (\ref{RN}), the selection rule for the exciton spin flip $m= -1 \to m= 1$ by $\Delta m =2$ reads 
\begin{equation} \label{Vexc}
	V^{++}_{f,i}(XX) \propto  [{\bm e}^* \cdot ({\bm o}_x - {\rm i} {\bm o}_y)][{\bm e}^0 \cdot ({\bm o}_x - {\rm i} {\bm o}_y)]\:.
\end{equation}

A direct evidence of participation of the resident exciton in the light scattering with the intermediate biexciton state can be an observation of a process in which the final state of the resident exciton has the angular momentum projection $m=0$. The selection rules for a change of the exciton spin component $m$ by 1 are similar to those for the single SFRS, Eq.~\eqref{CSCT1}. However, in this case the energy conservation law contains only half of the exciton Zeeman splitting, i.e., $\mu_B B (g_e+g_h)/2$, and it is determined by the half-sum of the electron and hole $g$ factors. Apparently, such the processes have been observed in CsPbBr$_3$ perovskite crystals \cite{Kalitucha2022}, along with single and double spin-flips of resident carriers.  

We have considered Raman scattering involving simultaneous spin flips of localized electron and hole or exciton localized as a whole. It is possible that an exciton excited by light emits or absorbs an acoustic phonon and changes the projection of its angular momentum by $\Delta m = \pm 2$. Such processes require separate consideration. 

\section{Spin-flip Raman scattering. Exciton polaritons} \label{excpol}
\subsection{Scattering of a free mechanical exciton by paramagnetic centers}  \label{SbA}
Before considering the scattering of exciton polaritons by paramagnetic centers, we will first solve an auxiliary problem of the scattering of a free mechanical exciton, without taking into account the longitudinal-transverse splitting of exciton states and the interaction with a transverse electromagnetic wave. The polarization of a triplet exciton with the wave vector ${\bm k}$ is described by the unit vector ${\bm c}_{\bm k}$ with the components $c_{{\bm k},x}, c_{{\bm k} ,y}, c_{{\bm k},z}$ which defines the exciton wave function similarly to Eq.~(\ref{exciton}).

As in the previous sections, the $z$ axis is chosen to be oriented along the external magnetic field ${\bm B}$. For definiteness, we assume the Lands\'e factor $g_e$ of localized resident electrons to be positive. We consider the scattering of an exciton by a paramagnetic center from the initial state $|{\bm k}, {\bm c}_{\bm k}; \downarrow_e \rangle$ (exciton with the wave vector ${\bm k}$ and spin state ${\bm c}_{\bm k}$ + localized electron with spin $\downarrow_e$) to the final state $|{\bm k}', {\bm c}_{{\bm k}'}; \uparrow_e \rangle$. For the contact interaction (\ref{exchange2}), the scattering matrix element has the form
\begin{eqnarray} 
	M^{({\rm exc})}_{ {\bm k}' {\bm c}_{{\bm k}'}, {\bm k} {\bm c}_{\bm k}} &\equiv& \langle {\bm k}', {\bm c}_{{\bm k}'}; \uparrow_e  |H_{\rm exch}  |{\bm k}, {\bm c}_{\bm k}; \downarrow_e  \rangle  \\  &=& \frac{\tilde{J}_{ee}  \Omega_0}{V}{\rm e}^{{\rm i} ({\bm k} - {\bm k}') {\bm r}_e^0} \Lambda({\bm c}_{{\bm k}'}, {\bm c}_{\bm k})\:,  \nonumber \\
	\Lambda({\bm c}_{{\bm k}'}, {\bm c}_{\bm k}) &=& - {\rm i} ( {\bm c}_{{\bm k}'}^* \times {\bm c}_{\bm k}) \cdot ({\bm o}_x - {\rm i} {\bm o}_y)\:. \label{Lambla}
\end{eqnarray}
In the derivation, the products $ka_B, k'a_B$, $a_B$ being the exciton Bohr radius, were assumed small compared to unity.

Using the golden rule of quantum mechanics, we can find the scattering probability per unit time
\begin{eqnarray} \label{wk'k}
&&	w_{ {\bm k}',{\bm k} }({\bm c}_{{\bm k}'}, {\bm c}_{\bm k}) = \frac{2 \pi}{\hbar} \frac{N_{p,e}}{V} (\tilde{J}_{ee}  \Omega_0)^2 \Lambda({\bm c}_{{\bm k}'}, {\bm c}_{\bm k}) |^2 \nonumber \\
&&\mbox{} \hspace{2.5 cm}	\times \delta(E_{k'} - E_k + \Delta_{Z,e})\:. 
\end{eqnarray}
Here $N_{p,e}$ is the concentration of paramagnetic centers with localized electrons, $E_k = \hbar^2 k^2/2 M$ is the kinetic energy of the exciton, $M$ is its translational mass, and $\Delta_{Z,e}$ is the Zeeman splitting of a localized electron $g_e \mu_B B$. Integrating over all directions of ${\bm k}'$ and summing over the exciton polarization states we find the inverse lifetime of the exciton ${\bm k}$ with respect to spin-flip scattering by paramagnetic centers
\begin{equation} \label{Tsf}
	\frac{1}{T_{sf}} = \frac{N_{p,e} M (\tilde{J}_{ee} \Omega_0)^2 k'_e}{\pi \hbar^3}\:,
\end{equation} 
where
\begin{equation} \label{k'}
	k'_e = \left( k^2 - \frac{2 M}{\hbar^2} \Delta_{Z,e} \right)^{1/2}\:.
\end{equation}
The expression (\ref{Tsf}) must be supplemented with the contribution from scattering by localized holes. As a result, the total spin relaxation time is given by
\begin{equation} \label{Tsfa}
	\frac{1}{T_{sf}} = \frac{M (N_{p,e} \tilde{J}^2_{ee} k'_e+ N_{p,h} \tilde{J}^2_{hh} k'_h) \Omega_0^2 }{\pi \hbar^3}\:,
\end{equation} 
where $N_{p,h}$ is the concentration of paramagnetic centers with localized holes,
\[
k'_h = \left( k^2 - \frac{2 M}{\hbar^2} \Delta_{Z,h} \right)^{1/2}\:,
\]
and $ \Delta_{Z,h} = g_h \mu_B B$.  For brevity, in what follows we will take into account only the spin-flip scattering of localized electrons. The generalization is performed in the same way as it is done in the formula (\ref{Tsfa}).

\subsection{Scattering of an exciton polariton by paramagnetic centers } \label{excpol2}
For simplicity, we neglect here the spatial dispersion and describe the medium by the permittivity
\begin{equation} \label{epsilon}
	\varepsilon(\omega) = \varepsilon_b \left( 1 + \frac{\omega_{LT}}{\omega_T - \omega} \right)\:,
\end{equation}
where $\omega_T$ is the resonant frequency of the mechanical exciton, $\omega_{LT}$ is the longitudinal-transverse splitting, $\varepsilon_b$ is the background permittivity, and the resonant region of the spectrum $|\omega_T - \omega| \ll \omega_T$ is considered. To further simplify the problem, we restrict ourselves to excitations in the frequency range below $\omega_T$. The light-exciton mixing, controlled by $\omega_{LT}$, leads to the transformation of the ``bare'' photon dispersion $c k/\sqrt{\varepsilon_b}$ into the polariton dispersion
\begin{equation} \label{dispersion}
	\omega_k = \frac{ck}{\sqrt{\varepsilon (\omega_k)}}\:,
\end{equation}
where $k = |{\bm k}|$, ${\bm k}$ is the polariton wave vector. As a rule, in the following we omit the index $k$ in the notation $\omega_k$.

In an isotropic medium, polaritons are transverse waves. We choose the unit polarization vectors of two degenerate polariton states with the wave vector ${\bm k}$ in the form
\[
{\bm c}_{{\bm k}1} = \frac{1}{k_{\perp} k}\left( k_zk_x, k_z k_y, - k_{\perp}^2 \right)\:,\:  {\bm c}_{{\bm k}2} = \left( -\frac{ k_y}{k_{\perp}}, \frac{k_x}{k_{\perp}}, 0 \right)\:,
\]
where $k^2_{\perp} = k_x^2 + k_y^2$.

The polariton annihilation operator $\alpha_{{\bm k}j}$ is related to the creation and annihilation operators $a^{\dag}_{{\bm k}j}, a_{{\bm k}j}$ of ``bare'' photons and similar operators $b^{\dag}_{{\bm k}j}, b_{{\bm k}j}$ for excitons by \cite{Hopfield}
\begin{eqnarray} \label{annih}
	\alpha_{{\bm k}j} =  w_{\bm k}  a_{{\bm k}j} + x_{\bm k}  b_{{\bm k}j} + y_{\bm k}  a^{\dag}_{-{\bm k}j} + z_{\bm k}  b^{\dag}_{-{\bm k}j}\:.
\end{eqnarray}
In the resonant frequency region, $\omega_T - \omega \ll \omega_T$, the square modulus of the coefficient $x_{\bm k}$, also called the strength function, is equal to
\[
|x_{\bm k}|^2 \equiv s(\omega) = \frac{\omega_{LT} \omega_T}{\omega_{LT} \omega_T + 2 (\omega_T - \omega)^2}\:.
\] 

When interacting with a paramagnetic center, the polariton scattering $\omega, {\bm k}, {\bm c}_{{\bm k}j} \to \omega', {\bm k}', {\bm c}_ {{\bm k}' j'}$ occurs due to the exciton component, and the squares of the modulus of the polariton and exciton scattering matrix elements are related by the relation, see, e.g., Ref. \cite{Yuldashev},
\begin{equation}
	|M^{({\rm pol})}_{{\bm k}'j', {\bm k}j }|^2 = s(\omega_{k'}) s(\omega_k) |M^{({\rm exc})}_{ {\bm k}' {\bm c}_{ {\bm k}' j' }, {\bm k} {\bm c}_{ {\bm k} j}} |^2\:.
\end{equation}
Furthermore, for the scattering probability we get instead of Eq. (\ref{wk'k})
\begin{eqnarray} \label{wk'k2}
&&	w_{ {\bm k}',{\bm k} }({\bm c}_{k'j'}, {\bm c}_{{\bm k}j}) = \frac{2 \pi}{\hbar} \frac{N_{p,e}}{V}
	J^2 s(\omega_{{\bm k}'}) s(\omega_k)    \\
	&&\mbox{} \hspace{1 cm}  \times |\Lambda({\bm c}_{{\bm k}'j'}, {\bm c}_{{\bm k}j}) |^2 \delta(\hbar \omega_{k'} - \hbar \omega_k + \Delta_{Z,e})\:.  \nonumber
\end{eqnarray}
The lifetime $T^{({\rm pol})}_{sf}$ for exciton-polariton scattering with spin flips of localized electrons is found by summing over all directions of ${\bm k}'$ and two states of transverse polarization ${\bm c} _{{\bm k}'j'}$ to result in
\begin{eqnarray} \label{Tsf2}
	&&\mbox{}\hspace{0.5 cm}\frac{1}{T^{({\rm pol})}_{sf}} = \frac{N_{p,e} (\tilde{J}_{ee} \Omega_0)^2 k^{\prime 2}}{\pi \hbar^2 v(\omega')} \Lambda^2
	\:,\\&&
	\Lambda^2 = \sum\limits_{j'=1,2}\overline{ |\Lambda({\bm c}_{{\bm k}'j'}, {\bm c}_{{\bm k}j}) |^2} = \frac23 (1 + c^{2}_{{\bm k}j;z})\:.
\end{eqnarray}  
Here the wave number $k'$ is defined as
\begin{equation} \label{kprime}
	k' = \frac{\omega'}{c} \sqrt{\varepsilon(\omega')}\:,\: \hbar \omega' = \hbar \omega - \Delta_{Z,e}\:,
\end{equation}
the group velocity is introduced by
\begin{equation} \label{vomega}
	v(\omega) = \frac{d \omega_k}{d k} =\frac{c}{n(\omega)} \left[ 1 + \frac{\omega_T \omega_{LT} }{2(\omega_T - \omega)(\omega_L - \omega)}\right]^{-1}\:,
\end{equation}
$n(\omega) = \sqrt{\varepsilon(\omega)}$ is the refraction index, and $\omega_L=\omega_T+ \omega_{LT}$ is the frequency of longitudinal exciton.

Further in this section, we use the notation: $\omega$ (instead of $\omega_0$ introduced in Sec. \ref{singleA} for the incident light) for the frequency of the polariton that did not undergo inelastic spin-flip scattering, $\omega'$ and $\omega''$ for the frequencies of the polariton after one and two inelastic scattering events, respectively.
\subsection{Polariton scattering efficiency for a short lifetime $\tau_0$}
Let us find the efficiency of spin-flip scattering under normal incidence of light and normal emission along the $z_l$ axis in the Voigt geometry (${\bm B} \parallel z \parallel x_l$) so that the polarization vectors of the secondary and primary light lie in the $(x_l, y_l)$ plane: the polarizer transmits light with a unit polarization vector ${\bm e}^0$, the analyzer detects secondary radiation with a polarization ${\bm e}$.

The spatial distribution of electromagnetic field inside the crystal is determined by the ratio between the time of elastic (isotropic) scattering of an exciton polariton ($\tau_p$), the polariton nonradiative lifetime ($\tau_0$), and the spin-flip scattering time $T^{ ({\rm pol})}_{sf}$, the latter defined by Eq.~ (\ref{Tsf2}). To find the elastic scattering time $\tau_p$, we model the nonmagnetic scattering potential by
\begin{equation} \label{Hint}
H_{\rm int} = \sum_{n} {\cal V}_0\ \delta({\bm R} - {\bm r}_n)\:,
\end{equation} 
where ${\cal V}_0$ is a constant coefficient, ${\bm r}_n$ is the position of the $n$-th point defect, ${\bm R}$ is the coordinate of the exciton center of mass. In this model we have according to \cite{Yuldashev}
\begin{equation}
	\frac{1}{\tau_p} = \frac{2}{3 \pi} \frac{{\cal V}_0^2 N_d k^2}{\hbar^2 v(\omega)}\:.
\end{equation}
where $N_d$ is the concentration of scattering defects. Note that the scattering by paramagnetic centers without spin flip of localized electrons or holes also contributes to $\tau_p^{-1}$. In order to simplify the description, we assume that the Zeeman splitting of an exciton polariton is small compared to the damping $\hbar \Gamma$, where $2 \Gamma$ is the sum of the inverse times $\tau_0^{-1}, \tau_p ^{-1}, T^{({\rm pol})-1}_{sf}$. This makes it possible to neglect the birefringence induced by the magnetic field ${\bm B}$.

In this subsection, we will analyze the case of a short lifetime, or the case 1,
\begin{equation} \label{Case1}
	\tau_0 \ll \tau_p, T^{({\rm pol})}_{sf}\:.
\end{equation} 
The limiting case of multiple elastic scattering, $\tau_p \ll \tau_0, T^{({\rm pol})}_{sf}$, (case 2) is considered in section \ref{SbD}.

Provided the condition (\ref{Case1}) is satisfied, it suffices to calculate scattering with a single spin flip, i.e., a first-order process whose probability per unit time is described by the formula (\ref{wk'k2}). The intensity of radiation singly scattered and emerging into vacuum in a solid angle $d \Omega$ is determined by
\begin{widetext}
\begin{eqnarray} \label{order1}
	d I(\omega', {\bm e}) &=& \frac{2 \pi}{\hbar} \frac{\hbar \omega' d \Omega}{n^2(\omega')} T(\omega')  s(\omega')s(\omega) (\tilde{J}^2_{ee} \Omega_0)^2 N_p |\Lambda({\bm e}, {\bm e}^0)|^2 \nonumber \\ &\times& \int\limits_0^{\infty} \frac{k^{\prime 2} dk'}{(2 \pi)^3} \delta(\hbar \omega_{k'} - \hbar \omega_k + \Delta_{Z,e}) \int\limits_0^{\infty} {\rm e}^{- \alpha(\omega') z_l} {\cal N}_{\bm k}(z_l) dz_l\:. 
\end{eqnarray}
\end{widetext}
where
\begin{equation} \label{calN}
	{\cal N}_{\bm k}(z_l) = \frac{T(\omega)}{\hbar \omega v(\omega)}F(\omega, {\bm e}^0) {\rm e}^{- \alpha(\omega) z_l}\:,
\end{equation}
$F(\omega, {\bm e}^0)$ is the energy flux of electromagnetic radiation (per unit area) incident on the sample from vacuum, $\alpha(\omega)$ is the absorption coefficient equal to $2 \Gamma s (\omega)/v(\omega)$, $T(\omega)$ is the light transmittance through the sample boundary, ${\cal N}_{\bm k}(z_l)$ is the density of polaritons with the wave vector ${\bm k}$ and polarization ${\bm e}^0$ at the point $z_l$. At normal incidence and normal luminescence, we have
\begin{eqnarray} \label{lambda}
&&	|\Lambda({\bm e}, {\bm e}^0)|^2 = |{\bm e}^* \times {\bm e}^0|^2\:, \\ 
&&	T(\omega) = \left\vert \frac{n(\omega) - 1}{n(\omega) + 1}\right\vert^2\:,\: T(\omega') = \left\vert \frac{n(\omega') - 1}{n(\omega') + 1}\right\vert^2\:. \nonumber
\end{eqnarray}
In the frequency range $(\omega_T - \omega)^2 \ll \omega_T \omega_{LT}$ of interest to us, the strength functions can be set equal to unity.

Substituting Eq.~(\ref{calN}) into Eq.~(\ref{order1}) and integrating over $z_l$ and $k'$, we come to
\begin{widetext}
\begin{eqnarray} \label{order1a}
	\frac{dI(\omega', {\bm e})}{d \Omega} &=&  \frac{T(\omega') T(\omega) }{4 \pi^2 \hbar^2 n^2(\omega')}  \frac{(\tilde{J}^2_{ee}\Omega_0)^2 N_{p,e} k^{\prime 2}}{v(\omega') v(\omega)} \frac{F(\omega, {\bm e}^0)}{\alpha(\omega') + \alpha(\omega)}|\Lambda({\bm e}, {\bm e}^0)|^2 \\ &=& \frac{T(\omega') T(\omega) }{4 \pi^2 \hbar^2} \left( \frac{\omega}{c} \right)^2 \frac{\tau_0 (\tilde{J}^2_{ee}\Omega_0)^2 N_{p,e} F(\omega, {\bm e}^0) }{v(\omega') + v(\omega)} |{\bm e}^* \times {\bm e}^0|^2\:. \nonumber
\end{eqnarray}
\end{widetext}
When deriving, we took into account that the ratio $\omega'/\omega$ can be replaced by 1.

If the Zeeman splitting of polariton states is comparable or exceeds $\hbar \Gamma$ one has to take into account magnetic-field induced birefringence leading to the difference of $v(\omega)$ and $T(\omega)$ for polaritons polarized parallel and perpendicular to the transverse magnetic field ${\bm B} \perp z_l$ in the Voigt geometry. Then Eq.~(\ref{order1a}) is valid for the linear polarizations ${\bm e}^0, {\bm e} \parallel$ or $\perp {\bm B}$ but overestimates the sensitivity of $dI(\omega', {\bm e})/d \Omega$ to the circular polarization. 
\subsection{The case of frequent elastic collisions $\tau_p \ll \tau_0, T^{({\rm pol})}_{sf}$} \label{SbD}

Exciton polaritons excited by the light of frequency $\omega$ and not undergone inelastic scattering have energy $\hbar \omega$. We denote their distribution function at the point $z_l$ as $m(\omega, {\bm \Omega},j;z_l)$, where ${\bm \Omega}$ is the unit vector ${\bm k} /k$ and $j$ is the polarization index. The distribution function and intensity of polaritons are related as $$I(\omega, {\bm \Omega}, j;z_l) = \hbar \omega v(\omega) m(\omega, {\bm \Omega},j;z_l )\:.$$
Taking into account the double degeneracy of transverse polaritons, their concentration is equal to
\[
N(\omega, z_l) = \sum\limits_{j=1,2} \int\limits_{4 \pi} d \Omega\ m(\omega, {\bm \Omega},j;z_l) \:.
\]

Strictly speaking, in the general case, instead of the function $m(\omega, {\bm \Omega},j)$, one should introduce the density matrix $m_{j'j}(\omega, {\bm \Omega})$. However, in the case 2 this matrix is diagonal in indices $j',j$, the distribution of polaritons over directions ${\bm \Omega}$ is uniform and they are not polarized: $m_{j'j}(\omega, {\bm \Omega}; z_l) = m(\omega, {\bm \Omega},j;z_l) \delta_{j'j} \equiv m(\omega;z_l) \delta_{j'j}$ and
$N(\omega, z_l) = 8 \pi m(\omega;z_l)$.

The concentration $N(\omega,z_l) \equiv N(z_l)$ satisfies the diffusion equation
\begin{equation}\label{diff}
	- D \frac{\partial^2 N(z_l)}{\partial z_l^2} + \frac{N(z_l)}{\tau} = 0\:,
\end{equation}
where $D$ is the diffusion coefficient $v^2(\omega) \tau_p/3$, the time $\tau$ is determined according to
\begin{equation}\label{tau}
	\frac{1}{\tau} = \frac{1}{\tau_0} + \frac{1}{\tau_{sf}}\:,
\end{equation}
$\tau_{sf}$ is the spin-flip scattering time, it is found from the expression (\ref{Tsf2}) for the inverse time $T_{sf}^{ ({\rm pol})}$ by averaging over the directions of ${\bm k}$ and polarizations ${\bm c}_{{\bm k}j}$
\begin{equation} \label{tausf}
	\frac{1}{\tau_{sf}} = \frac89 \frac{(\tilde{J}^2_{ee}\Omega_0)^2 N_{p,e} k^{\prime 2}}{\pi \hbar^2 v(\omega')} \:.
\end{equation}

The solution of Eq.~(\ref{diff}) is standard
\[
N(z_l) = N_0 {\rm e}^{- z_l/l}\:,
\]
where $l = \sqrt{D \tau}$ is the diffusion length. The concentration $N_0$ at the point $z_l=0$ is found from the boundary condition
\[
\left[ \frac{l}{\tau}  + v(\omega)T_{\rm eff}  \Delta \Omega \right] N_0 = \frac{T(\omega) F(\omega, {\bm e}^0)}{\hbar \omega}\:,
\]
where $\Delta \Omega$  is the solid angle limited by the angle of total internal reflection $\vartheta_{\rm cr}$, $T_{\rm eff}$ is the effective transmittance averaged over the angle of incidence $0 \leq \vartheta \leq \vartheta_{\rm cr}$ of a polariton to the inner boundary of the sample.

Let us introduce the effective time for the escape of polaritons into vacuum 
\[
\tau_{\rm vac} = \frac{l}{v(\omega)T_{\rm eff} \Delta \Omega}\:.
\]
For simplicity, we further assume the condition
\begin{equation} \label{condition}
\tau_0 \ll \tau_{\rm vac}\:. 
\end{equation}

\subsubsection{The first-order spin-flip scattering}

	\begin{figure}[h!]
	\centering
	\includegraphics[width=0.9\linewidth]{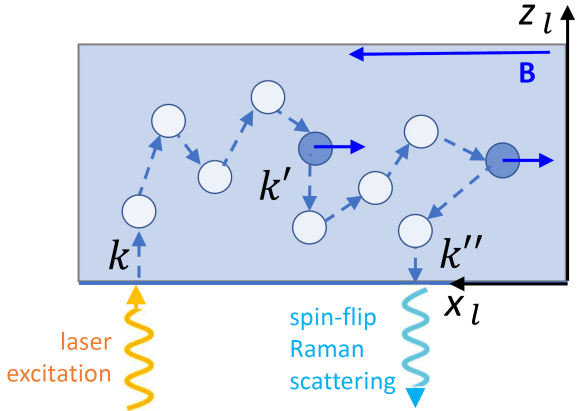}
	\caption{   SFRS by exciton polaritons in the regime of frequent elastic collisions, $\tau_p \ll \tau_0$. Schematic illustration of the multiple elastic scattering of the exciton polariton including two spin flips of the resident charge carriers.  	\label{Fig2}  } 
\end{figure}

After polaritons with energy $\hbar \omega$ experience spin-flip scattering by localized carriers, they continue elastic scattering from defects before leaving the resonant region within the time $\tau_0$ or being scattered by another paramagnetic center, Fig.~\ref{Fig2}.  We denote by a prime the parameters of polaritons with the energy $\hbar \omega' = \hbar \omega - \Delta_{Z,e}$. The diffusion equation for the concentration of such polaritons $N(\omega',z_l) \equiv N'(z_l)$ contains an inhomogeneous term in the right-hand side
\begin{equation} \label{diff2}
	- D' \frac{\partial^2 N'(x)}{\partial z_l^2} + \frac{N'(z_l)}{\tau'} =\frac{N(z_l)}{\tau_{sf}}\:.
\end{equation}
The general solution of this equation has the form
\begin{equation} \label{general}
	N'(z_l) = N'_0 {\rm e}^{- z_l/l'} +  N'_1 {\rm e}^{- z_l/l} \:,
\end{equation}
where $l' = \sqrt{D' \tau'}$,
\[
N'_1 =  \frac{N_0}{1 - (l'/l)^2}  \frac{\tau'}{\tau_{sf}}\:,
\]
and the density $N'_0$ is found from the boundary condition at $z_l=0$.
Under the condition (\ref{condition}) for polaritons with energy $\hbar \omega'$, the solution (\ref{general}) takes the form 
\begin{equation}
	N'(z_l) = N_0  \frac{\tau'}{\tau_{sf}} \frac{l (l {\rm e}^{- z_l/l}  - l' {\rm e}^{- z_l/l'})}{l^2 - l^{\prime 2} }\:. 
\end{equation}
The density at the sample inner surface equals to
\begin{equation} \label{Nomega1}
	N(\omega',0) =   \frac{\tau'}{\tau_{sf}} \frac{l N_0}{l + l'}\:,
\end{equation}
and, for the intensity of light outgoing along the normal, we have
\begin{equation} \label{dIdOmega}
	\frac{d I(\omega')}{d \Omega} = \hbar \omega' v(\omega')  N(\omega',0) \frac{T(\omega')}{n^2(\omega')} \:.
\end{equation}
\subsubsection{The second-order spin-flip scattering}
We denote by $N(\omega'', z_l) \equiv N''(z_l)$ the concentration of polaritons which participated in the double spin flip scattering by localized electrons ($\hbar \omega'' = \hbar \omega - 2 \Delta_{Z,e}$). The diffusion equation for them is similar to Eq.~(\ref{diff2}),
\begin{equation} \label{D2}
	- D'' \frac{\partial^2 N''(x)}{\partial z_l^2} + \frac{N''(z_l)}{\tau''} =\frac{N'(z_l)}{\tau'_{sf}}\:.
\end{equation}
The general solution and the boundary condition it satisfies have the form
\begin{eqnarray}
	N''(z_l) = N''_0 {\rm e}^{-z_l/l''} + N''_1{\rm e}^{-z_l/l} + N''_2{\rm e}^{-z_l/l'}  \:, \label{N2}\\ \frac{1}{\tau''} \left( l'' N''_0 + l N''_1 + l' N''_2\right) = \frac{l}{\tau'_{sf}} \frac{\tau'}{\tau_{sf}}N_0 \:, \hspace{0.5 cm}\mbox{} 
\end{eqnarray}
where $l'' = \sqrt{D'' \tau''}$. The coefficients $N''_1$ and $N''_2$ are defined according to
\begin{eqnarray}
	&&N''_1 = \frac{\tau''}{\tau'_{sf}} \frac{l^2}{l^2 - l^{\prime \prime 2}}N'_1\:,\\
	&&N''_2 = \frac{\tau''}{\tau'_{sf}} \frac{l^{\prime 2}}{l^{\prime 2} - l^{\prime \prime 2}}N'_0\:. \nonumber
\end{eqnarray}

Omitting the intermediate calculations we present the final expression for the concentration $N''$ at the boundary
\begin{equation} \label{N20}
	N''(0) =  \frac{h l}{l''}  \xi N_0 \:,
\end{equation}
where
\[
h = 1  +  \frac{1}{l^2 - l^{\prime 2}} \left( \frac{ l^{\prime 3}}{l' + l''} - \frac{ l^3}{l + l^{\prime \prime}} \right)\:,\: \xi = \frac{\tau'' \tau'}{\tau'_{sf} \tau_{sf}}\:.
\]
According to Eqs.~(\ref{Nomega1}), (\ref{dIdOmega}), and (\ref{N20}), the ratio of the second and first order intensities is
\begin{equation} \label{21}
	\frac{d I(\omega'')/ d \Omega}{d I(\omega') / d \Omega} = h \frac{\tau''}{\tau'_{sf}} \ \frac{T(\omega'')n^2(\omega')}{T(\omega')n^2(\omega'')} \ \frac{l + l'}{l''}  \:.
\end{equation}
Thus, in the regime of frequent collisions with the comparable lifetimes $\tau_0$ and $\tau'_{sf}$, the intensities of single and double scattering are comparable as well.

It should be mentioned that in the case where the Zeeman splitting of polariton states are large, the diffusion equations remain also valid, however comprise renormalized times  $\tau_0, \tau_{sf}, \tau', \tau_{sf}', \tau''$ and the diffusion coefficients.

\section{Discussion: Analysis of the polarization dependence of the SFRS}	\label{VI}
We start the analysis from a single spin-flip of a localized electron or hole due to the interaction with a photoexcited localized exciton. 

At normal light incidence in the positive direction of the $z_l$ axis of the laboratory frame (see Fig. \ref{Fig1}) and a backward registration of scattered light, $\bar z_l$, the polarization of light has components only along the axes $x_l, y_l$. In this case, the vector product in Eq.~(\ref{CSCT1}) has a single nonzero projection onto the $z_l$ axis. The projections ${\bm o}_x \pm i {\bm o}_y$ on the axis $z_l$ coincide, the square of their modulus can be found from Fig. \ref{Fig1}(a) as the square of the projection of the unit vector ${\bm b}$ on the plane $(x_l,y_l)$, equal to $ |({\bm o}_x \pm i {\bm o}_y)_{z_l}|^2 = 1- b_{z_l}^2 = 1-\cos^2 \theta$. Thus,  we obtain from Eq.~(\ref{CSCT1}) that the scattering intensity depends on the light polarization and the direction of the magnetic field as 
from (\ref{CSCT1}) we obtain that the intensity of the process depends on the polarization of the light and the direction of the magnetic field as 
	\begin{equation} \label{int1}
	I_{+(-)}^{(1e)}, I_{+(-)}^{(1h)}  \propto |{\bm e}^* \times {\bm e}^0|^2 \sin^2 \theta \, .
	\end{equation}
	
Hence, the single scattering process is forbidden in the Faraday geometry [$\theta = 0, \pi$, Fig. \ref{Fig1}(b)] and characterized by strict selection rules in the general geometry of Fig. \ref{Fig1}(a): Scattering in the crossed linear (${\bm e} \perp {\bm e}^0$) and co-circular ($\sigma_+, \sigma_+$~or~$\sigma_-, \sigma_-$) configurations is allowed and occurs with the equal probability proportional to $\sin^2{\theta}$, while in co-linear (${\bm e} \parallel {\bm e}^0$) and crossed circular ($\sigma_-, \sigma_+$~or~$\sigma_-, \sigma_+$) configurations the scattering is prohibited. Note that at normal incidence, the $\sigma_{\pm}$ circular polarization orts are defined according to $({\bm o}_{x_l} \pm {\rm i} {\bm o}_{y_l})/\sqrt{2}$, where ${\bm o}_{x_l}, {\bm o}_{y_l}$ are unit vectors along the laboratory axes $x_l$ and $y_l$. Therefore, the scattered light of the $\sigma_+$ polarization ${\bm e} =({\bm o}_{x_l} + {\rm i} {\bm o}_{y_l})/\sqrt{2} $ is right-handed if it is scattered forward and left-handed if it is scattered backwards. An important property of the single SFRS is its independence of the magnitude of the exchange splitting between the bright and dark exciton.

The same selection rules \eqref{int1} characterize the exciton-polariton scattering with a single spin flip of the resident carrier in the case  of short lifetime $\tau_0$. In contrast, in the case of frequent elastic collisions, Fig. \ref{Fig2}, information about the initial polarization of the exciton polariton is lost, and the SFRS is not polarized. Thus, a measurement of the polarization of secondary light in the polariton region allows one to estimated the ratio between the times $\tau_0$ and $\tau_p$.

Note that the $(\sigma_+, \sigma_+)$ or $(\sigma_-, \sigma_-)$ single scattering becomes possible if one takes account of the Larmor precession of the unpaired particle in the three-particle complex $2e+h$ or $e + 2h$. The effect is governed by the ratio $|g_{e,h}| \mu_B B / \hbar \Gamma$ which is here assumed to be small, Eq.~(\ref{gehG}).
	
Let us proceed to the double scattering. It follows from the selection rules \eqref{RN}, \eqref{Vexc}  that the light scattering intensity because of simultaneous spin flips of the resident electron and hole or the resident exciton depends on the light polarization and the magnetic-field direction as follows
	\begin{eqnarray}  \label{intensity3}
&&	I_{++}^{(1e,1h)}, I_{++}(XX) \propto |V^{++}_{f,i}|^2   \\ && \propto \left( 1 - |{\bm e} {\bm b}|^2 - {\bm \kappa} {\bm b}\right) \left( 1 - |{\bm e}^0 {\bm b}|^2 + {\bm \kappa}_0 {\bm b} \right), \nonumber \\
&&	I_{--}^{(1e,1h)}, I_{--}(XX) \propto |V^{--}_{f,i}|^2 \\ && \propto \left( 1 - |{\bm e} {\bm b}|^2 + {\bm \kappa} {\bm b}\right) \left( 1 - |{\bm e}^0 {\bm b}|^2 - {\bm \kappa}_0 {\bm b} \right), \nonumber
	\end{eqnarray}
where
	\begin{equation} \label{kappa}
		{\bm \kappa} = {\rm i} ({\bm e} \times {\bm e}^*) = P_{\rm circ}\frac{\bm k}{k}\:,
	\end{equation}	 
$P_{\rm circ}$ and ${\bm k}$ are  degree of circular polarization and the wave vector of scattered light; ${\bm \kappa}_0$ is defined in the same way but for incident light. 
	
The same selection rules hold for the processes $I_{++}^{(2e (2h))}$ and $I_{--}^{(2e(2h))}$
which allows us, for the sake of brevity, to use the notation $I_{++}$ and $I_{--}$ for the scattering processes with the total-spin changes $\Delta m= +2$ or $\Delta m= -2$, respectively. 
 
One can see from Eqs.~(\ref{intensity3}) that in whatever geometry the scattering signal is forbidden if either the incident or scattered light is linearly polarized along the magnetic field and otherwise allowed. In the Faraday geometry ($\theta = 0, \pi$) and the Voigt geometry ($\theta = \pi/2$),  the polarization dependencies are represented by
\begin{eqnarray} \label{perplong}
&& I_{++}(\theta = 0),   I_{--}(\theta = \pi)   \propto (1 - \kappa_{z_l}) (1 + \kappa_{z_l}^0) \:, \\ 
 &&    I_{++}(\theta = \pi), I_{--}(\theta = 0)  \propto (1 + \kappa_{z_l}) (1 - \kappa_{z_l}^0) \:, \nonumber\\ 
  && I_{++}(\theta = \pi/2), I_{--}(\theta = \pi/2) \propto |{\bm e}_\perp|^2 |{\bm e}_\perp^0|^2\:, \nonumber
\end{eqnarray}
where ${\bm e}_\perp$ is the polarization vector component transverse to the magnetic field. In the Voigt geometry for circularly polarization of any handedness, $|{\bm e}_\perp|^2 = |e_z|^2=1/2$.  The polarization dependence (\ref{perplong}) for specific configurations at normal incidence is presented in Table \ref{TMDPAR}. The sums of numbers over the four linear and four circular configurations coincide and are four times larger in the Faraday geometry than in the Voigt geometry. 
 \begin{table*}[hpt]
 		\caption{Polarization dependence of the intensity $I_{++}$ for double SFRS in the Faraday ($\theta = 0, \pi$) and Voigt geometry ($\theta = \pi/2$). The double symbols in the row Configuration denote the polarizations of  incident  and scattered light, respectively; in the Voigt geometry $\parallel$ and $\perp$ are linear polarizations along and perpendicular to the external magnetic field, and in the Faraday geometry the symbol pairs $(\parallel, \parallel), (\parallel, \perp)$ etc. mean parallel and crossed linear polarizations of arbitrary azimuthal orientation. The numbers given in the Table are the values of the right-hand side expressions in Eqs.~(\ref{perplong}).}
	\begin{center}
		\begin{tabular}{|c|c|c|c|c|c|c|}
			\hline
			Configuration & $\parallel, \parallel$ & $\perp, \perp$ & $\perp, \parallel$ or $\parallel, \perp$ & $\sigma_+, \sigma_+$ or $\sigma_-, \sigma_-$& $\sigma_-, \sigma_+$ & $\sigma_+, \sigma_-$\\ \hline $\theta = 0$ &1&1&1&0&0&4 \\ \hline $\theta = \pi$ & 1&1&1&0&4&0\\
			\hline $\theta = \pi/2$ &0&1&0&1/4&1/4 &1/4\\ \hline
		\end{tabular}
	\end{center}
	\label{TMDPAR}
\end{table*}
\begin{table*}[hpt]
	\caption{Polarization dependence of the $+-$ double scattering intensity for the particular cases ${\rm i}, {\rm ii}, {\rm iii}$ and ${\rm iv}$ in the Faraday ($\theta = 0, \pi$) and Voigt ($\theta = \pi/2$) geometries. The double symbols denote the polarization configurations of incident and scattered light, the same as in Table \ref{TMDPAR}. }
	\begin{center}
		\begin{tabular}{|c|c|c|c|c|c|}
			\hline
			Configuration & $\parallel, \parallel$ & $\perp, \perp$ & $\perp, \parallel$ or $\parallel, \perp$ & $\sigma_+, \sigma_+$ or $\sigma_-, \sigma_-$& $\sigma_-, \sigma_+$ or $\sigma_+, \sigma_-$\\ \hline 
			$I^{({\rm i})}_{+-}$; $\theta = 0$ or $\pi$ &$0$&$0$&0&$0$&0\\
			\hline $I^{({\rm i})}_{+-}$; $\theta = \pi/2 \hspace{4 mm}$ &1&$0$&0&$1/4$&1/4\\
			\hline \hline $I^{({\rm ii})}_{+-}$; $\theta = 0$ or $\pi$ &$1$&$1$&0&$0$&1\\
			\hline $I^{({\rm ii})}_{+-}$; $\theta = \pi/2 \hspace{4 mm}$ &0&$1$&0&$1/4$&1/4\\ 
			\hline \hline $I^{({\rm iii})}_{+-}$; $\theta = 0$ or $\pi$ &$2$&$2$&0&$0$&2\\
			\hline $I^{({\rm iii})}_{+-}$; $\theta = \pi/2 \hspace{4 mm}$ &4&$2$&0&$1/2$&5/2 \\
			\hline \hline $I^{({\rm iv})}_{+-}$; $\theta = 0$ or $\pi$ &$1$&$1$&0&$0$&1\\
			\hline $I^{({\rm iv})}_{+-}$; $\theta = \pi/2 \hspace{4 mm}$ &4&$1$&0&$1/4$&9/4\\ 
			\hline
		\end{tabular}
	\end{center}
	\label{R0-+}
\end{table*}

Now we turn to the polarization dependence of the $+-$ scattering where the total spin of the carriers does not change, $\Delta m=0$. According to Eqs. (\ref{R1(h-e+)}), (\ref{R0(h-e+)}), the matrix elements of such processes contain combinations of the invariants $({\bm e}^* \cdot {\bm e}^0)$ and $({\bm e}^* \cdot {\bm b}) ( {\bm e}^0 \cdot {\bm b})$. The first invariant is independent of the magnetic field direction and has a maximum for co-linear and crossed circular configurations, whereas the second invariant vanishes in the Faraday geometry and, moreover, in the Voigt geometry if at least one of the vectors ${\bm e}, {\bm e}^0$ is perpendicular to the magnetic field. The partial contribution of the two invariants to the scattering intensity depends on the ratio between the bright-dark splitting $\Delta_{\rm exc} = |E_1-E_0|$ and the broadening of the intermediate states ``exciton plus resident carriers'', $\hbar \Gamma_0, \hbar \Gamma_1$. Four special cases can be distinguished as follows:

(i)  $\Delta_{\rm exc}  \ll \hbar \Gamma_0$$\ \approx \hbar \Gamma_1$ $\Rightarrow I^{({\rm i})}_{+-} \propto |R_1^{(+-)} + R_0^{(+-)}|^2$ $\mbox{} \hspace{5 cm} = |{\bm e} \cdot {\bm b}|^2|{\bm e}^0 \cdot {\bm b}|^2$\:,

(ii)  $\Delta_{\rm exc}   \approx \hbar \Gamma_0 \ll \hbar \Gamma_1$ $\Rightarrow$ 
$I^{ ( {\rm ii} ) }_{+-} \propto |R_0^{(+-)}|^2$\:,

(iii) $\hbar \Gamma_0 \ll \Delta_{\rm exc}   \approx \hbar \Gamma_1$ $\Rightarrow$ $ I^{({\rm iii})}_{+-}  \propto |R_1^{(+-)}|^2 + |R_0^{(+-)}|^2$,

(iv)  $\Delta_{\rm exc} \gg \hbar \Gamma_1, \hbar \Gamma_0$ $\Rightarrow$ $ I^{({\rm iv})}_{+-} \propto |R_1^{(+-)}|^2$\:.

Table \ref{R0-+} presents the relative intensities $I^{(\alpha)}_{+-}$ ($\alpha ={\rm i}, {\rm ii}, {\rm iii}, {\rm iv}$) for the $+-$ type of scattering in four limiting cases in the Faraday and Voigt geometries at normal incidence.Note that $I^{({\rm iii})}_{+-} = I^{({\rm ii})}_{+-} + I^{({\rm iv})}_{+-}$. As seen from the Table, the strongest signal is achieved in the Voigt geometry in the co-linear polarizations parallel to the magnetic field in the case of a large exchange splitting in the exciton.

As for the double spin-flip scattering of resident carriers involving exciton polaritons, such scattering is unlikely in the case of small exciton polariton lifetime (case 1). In the regime of frequent collisions (case 2, Fig. \ref{Fig2}), the secondary emission, both in single- and double spin-flip scattering, is not polarized or weakly polarized.  

The spherical invariant (\ref{CSCT1}) is applicable to an isotropic medium where the bright-exciton level is threefold degenerate. This model is valid for the cubic phase of a perovskite crystal.  When the symmetry is lowered to tetragonal, the triplet exciton degeneracy is partially removed and the scattering selection rules may be modified. In the case of a large energy splitting between exciton sublevels with the angular-momentum projection $\pm 1$ and $0$ on the tetragonal axis $c$, only the exciton states $\pm 1$ are important in the scattering process.  Then the scattering intensities are proportional to $|({\bm e}^*\times{\bm e}^0) \cdot {\bm c}|^2 \sin^2 \theta_{bc}$, where ${\bm c}$ is the ort along the $c$ axis and $\theta_{bc}$ is the angle between ${\bm c}$ and the magnetic field ${\bm B}$. In addition, the tetragonal phase allows for anisotropy of electron and hole $g$ factors \cite{Kirstein2022,Kalitucha2022}, and the description of SFRS gets closer to that for the spin reversal of resident electrons in CdSe nanoplatelets \cite{Rodina2020,Kudlacik2020}. The anisotropy also changes a value of the Stokes shift for the scattering involving the transition of the resident exciton from the state $m = -1$ to $m=0$. In Ref.~\cite{Kalitucha2022}, a splitting between exciton sublevels $\pm 1$ and $0$ at zero magnetic field has recently been demonstrated on a strained perovskite crystal.
\section{Conclusion} \label{VII}
Motivated by recent studies of semiconductor perovskites, we have developed a detailed theory of single and double spin-flip Raman scattering (SFRS) in bulk perovskite crystals. To gain an insight into the problem we have examined possible mechanisms of the scattering phenomenon, namely, (i) spin reversals of resident localized electrons and holes separated from each other in space and (ii) spin flips of nonequilibrium  excitons localized as a whole.  We have considered different types of resonant intermediate states involved in the scattering process: 
the complexes ``localized exciton + localized resident electron and hole'' and biexciton as well as free exciton polariton. To our knowledge, the participation of exciton polaritons is a new mechanism of SFRS, and this paper generalizes the theory of exciton polariton transport to take into account the exchange interaction of the electron-hole component of a polariton with localized charge carriers during the scattering. The results are presented in the form allowing one for the symmetry analysis, the scattering intensities are expressed in terms of the symmetry invariants applicable for the cubic phase of perovskites. 

Possibilities of future work can be divided into three directions. Firstly, the developed theory of SRFS by exciton polaritons can be extended from the frequency range $\omega < \omega_T$ to the whole resonance region with allowance for the spatial dispersion and additional light waves. Secondly, the theory may be modified to account for the effects of anisotropy in tetragonal and orthorhombic perovskites. Thirdly, the theory allows for an extension to third- and higher-order spin-flip scattering processes.

\section*{Acknowledgments}
 We thank D. R. Yakovlev, V. F. Sapega and I. V. Kalitukha for fruitful discussions. This work was funded by the Russian Foundation for Basic Research (Project 19-52-12064).

\end{document}